\documentclass[conference]{IEEEtran}
\IEEEoverridecommandlockouts
\usepackage{cite}
\usepackage{amsfonts} 
\usepackage{amsmath,amssymb,amsfonts}
\usepackage{algorithmic}
\usepackage[linesnumbered,ruled,vlined]{algorithm2e}
\usepackage{graphicx}
\usepackage{textcomp}
\usepackage{xcolor}
\usepackage{graphicx}
\usepackage{caption}
\usepackage{subcaption}
\usepackage{makecell}
\usepackage{bm}

\SetKwInput{KwInput}{Input}               
\SetKwInput{KwOutput}{Output}  

\graphicspath{{./Figures/}}

\def\BibTeX{{\rm B\kern-.05em{\sc i\kern-.025em b}\kern-.08em
    T\kern-.1667em\lower.7ex\hbox{E}\kern-.125emX}}
\begin{document}

\title{Practical Adversarial Attacks Against AI-Driven Power Allocation in a Distributed MIMO Network} 

\author{\IEEEauthorblockN{Ömer Faruk Tuna}
\IEEEauthorblockA{\textit{Ericsson Research} \\
Istanbul, Turkey \\
omer.tuna@ericsson.com}
\and
\IEEEauthorblockN{Fehmi Emre Kadan}
\IEEEauthorblockA{\textit{Ericsson Research} \\
Istanbul, Turkey \\
fehmi.emre.kadan@ericsson.com}
\and
\IEEEauthorblockN{Leyli Karaçay}
\IEEEauthorblockA{\textit{Ericsson Research} \\
Istanbul, Turkey \\
leyli.karacay@ericsson.com}
\and
}

\maketitle

\begin{abstract}
In distributed multiple-input multiple-output (D-MIMO) networks, power control is crucial to optimize the spectral efficiencies of users and max-min fairness~(MMF) power control is a commonly used strategy as it satisfies uniform quality-of-service to all users. The optimal solution of MMF power control requires high complexity operations and hence deep neural network based artificial intelligence (AI) solutions are proposed to decrease the complexity. Although quite accurate models can be achieved by using AI, these models have some intrinsic vulnerabilities against adversarial attacks where carefully crafted perturbations are applied to the input of the AI model. In this work, we show that threats against the target AI model which might be originated from malicious users or radio units can substantially decrease the network performance by applying a successful adversarial sample, even in the most constrained circumstances. We also demonstrate that the risk associated with these kinds of adversarial attacks is higher than the conventional attack threats. Detailed simulations reveal the effectiveness of adversarial attacks and the necessity of smart defense techniques.
\end{abstract}


\section{Introduction}

Deep learning is expected to be an important enabler for many wireless communication challenges in 6G. Deep neural networks (DNNs) are being proposed to handle a wide range of wireless communication tasks including encoding/decoding operations, spectrum sensing and RF signal classification. Power allocation for D-MIMO networks is one of the challenging tasks that we see the utilization of DNNs \cite{hexax}.

D-MIMO is a new network type considered for 6G communication systems where many radio units (RUs) are geographically distributed in a region to increase the coverage and reliability. The results obtained in \cite{demir2021foundations} show the benefits of D-MIMO compared to traditional uncoordinated small-cells which are used in all previous generations and collocated massive MIMO which is currently used in 5G systems. 

In D-MIMO, a precoding is applied at each RU to spatially focus the jointly transmitted signal on the desired user equipment (UE). There are different precoding techniques studied in the literature. Maximal ratio transmission (MRT) is one of the most attractive techniques where each RU applies a local precoding. To mitigate the multi-user interference, some power control techniques are applied together with precoding. According to the slowly changing channel statistics, RUs allocate their power among users to optimize the system's performance. Max-min fairness (MMF) power control aims to maximize the minimum spectral efficiency (SE) of users, and it is shown that the result satisfies uniform quality-of-service to all users. Future generation communication systems are required to satisfy uniformly good services to all users in an environment with ubiquitous demand, making MMF highly attractive for power control. On the other hand, the analytical solution of MMF requires high complexity operations. 

To overcome the complexity problem, Bashar et al. \cite{bashar2020exploit} propose deep learning methods to approximate the analytical solution using a low-complexity artificial intelligence (AI) model. Although AI models can accurately approximate the analytical result, it is known that they have some intrinsic properties which make them vulnerable to adversarial attacks~\cite{szegedy, Tuna2022}. Adversarial machine learning has lately attracted significant attention in wireless security domain as DNNs have become more commonly deployed. 

In this study, we investigate the potential effects of adversarial attacks targeting AI-driven power control systems in D-MIMO. We explain the main constraints of the adversary resulting from the distributed nature of wireless domain and focus only on the possible practical scenarios to observe the severity of adversarial attack threats. We work on attacks based on universal adversarial perturbation (UAP) which are not based on the exact input samples of the DNN but instead generalize the characteristics of a randomly selected subset of input samples. We propose a novel modified UAP (m-UAP) technique that crafts a specific perturbation for each input where there is only a partial knowledge about some of input entries. By performing several numerical simulations, we show that m-UAP with a surrogate AI model has much larger disruptive effect than standard Gaussian noise perturbation. To the best of our knowledge, this is the most practical work on the security of AI-driven power allocation in D-MIMO.

The organization of the paper is as follows. Section II describes the system model. In Section III, we present possible adversarial attacks. Section IV includes detailed numerical simulations, and Section V concludes the paper.

\section{System Model}
We consider a D-MIMO network with $M$ single antenna RUs which are connected to a central processor (CP) via wired fronthaul links. We assume that $K$ single antenna UEs are jointly served by all RUs in a given time/frequency resource block. In Fig. 1, we present an example D-MIMO network with 16 RUs, 4 UEs, and a CP.

\begin{figure}[ht]
    \centering
    \includegraphics[width=0.35\textwidth]{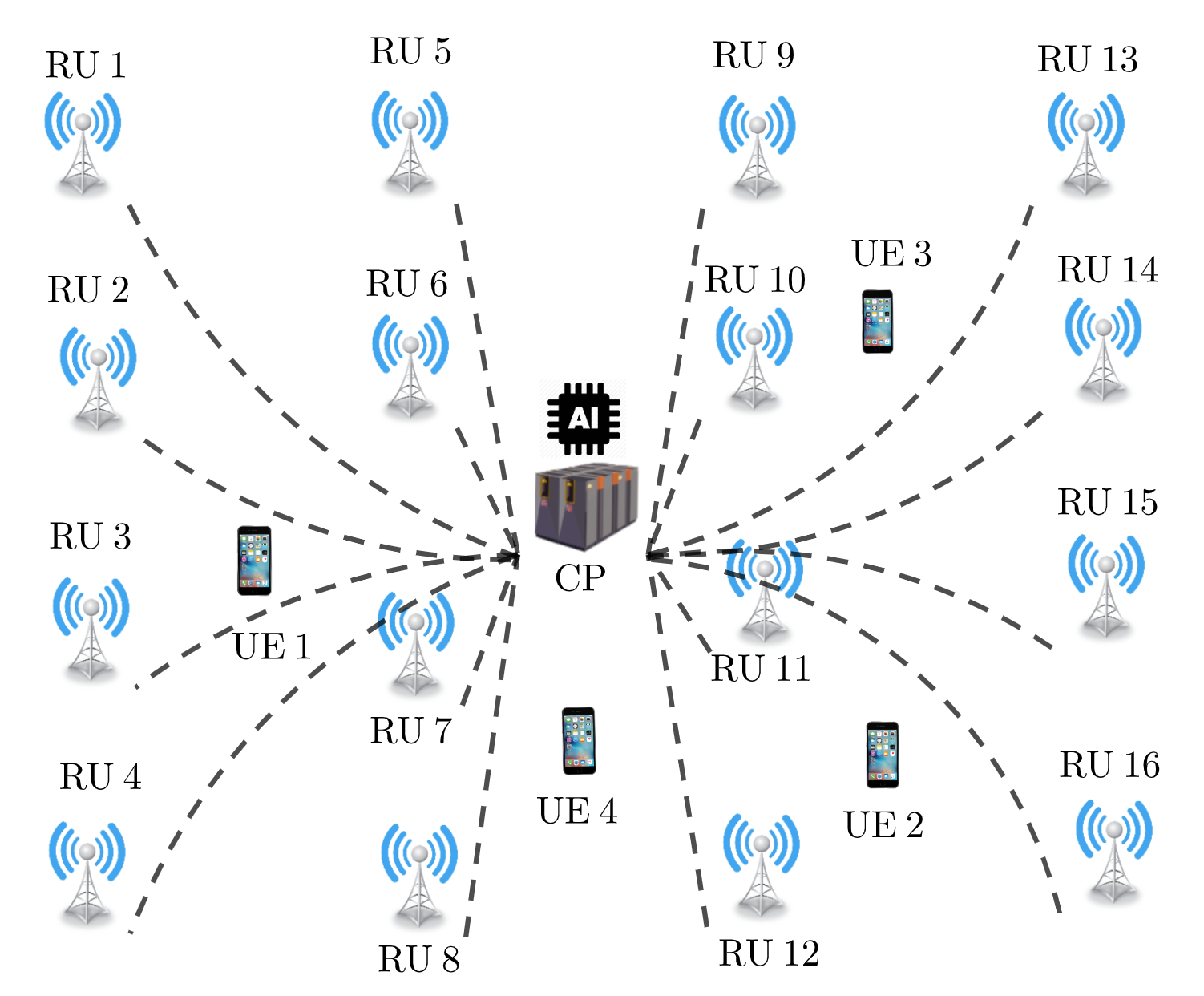}
    \caption{An example D-MIMO network.}
    \label{fig1}
\end{figure}

RUs apply MRT precoding where conjugate of the channel vectors are used to precode the user data. In this case, the transmitted signal from $m$-th RU can be written as
\begin{equation}
    x_m = \displaystyle\sum_{k=1}^K w_{m,k}s_k
\end{equation}
where $w_{m,k}=\sqrt{P_t\eta_{m,k}}h_{m,k}^{*}$ is the precoding vector of the $m$-th RU for the $k$-th user, $s_k$ is the information symbol of the $k$-th user satisfying $\mathbb{E}[|s_k|^2]=1$ for all $k$, $P_t$ is the maximum total transmit power of each RU, $\eta_{m,k}$ is the power control coefficient, and $h_{m,k}$ is the channel coefficient between RU $m$ and UE $k$. To satisfy transmit power constraints, we need
\begin{equation}
    \mathbb{E}[|x_m|^2] = \displaystyle\sum_{k=1}^K \mathbb{E}[|w_{m,k}|^2] = \displaystyle\sum_{k=1}^K \eta_{m,k}\beta_{m,k} \leq P_t, \quad \forall m
\end{equation}
where $\beta_{m,k}$ is the large-scale fading coefficient of the channel between RU $m$ and UE $k$, i.e., $\mathbb{E}[|h_{m,k}|^2]=\beta_{m,k}$. 

The received signal for the $k$-th user can be written as
\begin{equation}
    r_k = \displaystyle\sum_{m=1}^M h_{m,k} \displaystyle\sum_{\ell=1}^K w_{m,\ell}s_{\ell} + z_k
\end{equation}
where $r_k$ is the received signal and $z_k\sim \mathcal{C}\mathcal{N}(0, \sigma_k^2)$ is the receiver noise for the $k$-th user. Here, we assume that the channel coefficients $h_{m,k}$ are perfectly known at RU side.  The channel estimation can be performed using uplink pilots in time-division duplex (TDD) mode and when the channel coherence time is large enough, the channel coefficients can be estimated accurately. 

We assume that the users only have statistical channel knowledge, and they know the mean value of the effective channels. This assumption is widely used in the literature \cite{demir2021foundations},\cite{ngo2017cell}, \cite{nayebi2017precoding}. When the number of RUs is large, by means of channel hardening, the effective channels become nearly deterministic and hence, statistical knowledge is sufficient. Under this assumption, the received signal of the $k$-th user is given by
\begin{equation}
    r_k = r_{k, \text{desired}}+r_{k, \text{mismatch}}+r_{k, \text{interference}}+r_{k, \text{noise}}
\end{equation}
where
\begin{align}
    r_{k, \text{desired}} &= \displaystyle\sum_{m=1}^M\sqrt{P_t\eta_{m,k}}\mathbb{E}[|h_{m,k}|^2] \\
    r_{k, \text{mismatch}} &= \displaystyle\sum_{m=1}^M\sqrt{P_t\eta_{m,k}}|h_{m,k}|^2 - \sqrt{P_t\eta_{m,k}}\mathbb{E}[|h_{m,k}|^2] \notag \\
    r_{k, \text{interference}} &= \displaystyle\sum_{\ell \neq k}\displaystyle\sum_{m=1}^M\sqrt{P_t\eta_{m,\ell}}h_{m,k}h_{m,\ell}^{*}, \: \: r_{k, \text{noise}} = z_k. \notag
\end{align}
Here $r_{k, \text{desired}}$ is the desired signal part, $r_{k, \text{mismatch}}$ is the part involving the uncertainty of the effective channel, $r_{k, \text{interference}}$ includes multi-user interference, and $r_{k, \text{noise}}$ is the noise part. 

Using the approach given in \cite{demir2021foundations} and \cite{medard2000effect}, achievable user SEs can be given as \cite{nayebi2017precoding}
\begin{equation}
\begin{aligned}
\text{SINR}_k &= \dfrac{P_t\left(\displaystyle\sum_{m=1}^M \sqrt{\eta_{m,k}}\beta_{m,k}\right)^2}{P_t\displaystyle\sum_{\ell=1}^K\displaystyle\sum_{m=1}^M \eta_{m,\ell}\beta_{m,\ell}\beta_{m,k}+\sigma_k^2} \\
\text{SE}_k &= \log_2(1+\text{SINR}_k), \quad \forall k.
\end{aligned}
\end{equation}
Energy efficiency (EE) is another important parameter of networks which is aimed to be as large as possible. It is generally defined as \cite{ngo2017cell}
\begin{equation}
\text{EE} = \dfrac{\displaystyle\sum_{k=1}^K \text{BW}_k\cdot \text{SE}_k}{P_t\displaystyle\sum_{m=1}^M \displaystyle\sum_{k=1}^K\eta_{m,k}\beta_{m,k}}
\end{equation}
where EE is the average energy efficiency, $\text{BW}_k$ is the frequency bandwidth allocated for the $k$-th user. Here we ignore the energy consumption related to the fronthaul links and the other parts of the network, and focus on the energy efficiency of the access link only. EE shows the average number of bits transmitted per consumed energy unit. 

To optimize the power allocation, we consider MMF problem (P0) which can be defined as
\begin{equation}
(\text{P}0) \: \: \underset{\eta_{m,k}}{\text{max}} \: \min_k \text{SE}_k \quad \text{s.t.} \quad \displaystyle\sum_{k=1}^K \eta_{m,k}\beta_{m,k} \leq 1, \: \: \forall m.
\end{equation}
The optimal solution of the problem (P0) requires iterative quadratic programming which is of high complexity. To find an approximate and low-complexity solution deep learning method is proposed in the literature \cite{bashar2020exploit}. In this study, we assume that a DNN model is implemented at CP. The model is trained using an input data consisting of $\beta_{m,k}$'s and the output $\eta_{m,k}$'s used for training are calculated by solving (P0) analytically and in an offline fashion. The model details and the channel models to produce $\beta_{m,k}$'s are described in section \ref{simulation_result}.

\section{Adversarial Attacks in D-MIMO}

\subsection{Adversarial Attacks}

The goal of the adversarial attacks is to craft a perturbation $\delta$ under given constraints ($\delta<\epsilon$) which yields to an incorrect prediction as $f(x+\delta)=y^{adv}$ which differs from a prediction on a clean sample $f(x)=y$. The success criteria of the attack might change depending on the type of task. For a classification task, the attack can be considered successful if the model predicts a class other than the actual class. However, for regression tasks, many different objectives can be considered. In this study, our objective is to minimize user SEs. 

Fast Gradient-Sign Method (FGSM) \cite{goodfellow2015explaining} is one of the most popular adversarial attacks in literature which utilizes the derivative of the model's loss function with respect to the input to determine in which direction the feature values of the input vector should be changed to minimize the loss function of the model. Once this direction is extracted, it changes all features simultaneously in the opposite direction to maximize the loss. Later, Kurakin et al. \cite{kurakin2017adversarial} proposed a small but effective improvement to the FGSM, known as Basic Iterative Method (BIM). In this approach, rather than taking only one step of size $\epsilon$ in the gradient sign's direction, the attacker takes several but smaller steps $\alpha$, and use the given $\epsilon$ value to clip the result. Crafting adversarial samples under $L_{\infty}$ norm for BIM attack is given by (\ref{eq:bim}).
\begin{equation}
\begin{aligned}
\mathbf{x}_1^* & = \mathbf{x} \quad \text{and for all} \: \: i=1, 2, \ldots, i_{\text{max}} \\
\mathbf{x}_{i+1}^* & = \text{Clip}_{\epsilon} \{ \mathbf{x}_i^* + \alpha \cdot \text{sign} \left( \nabla_{\mathbf{x}} \left(J(\mathbf{x}_i^*, \bm{\theta}) \right)\right) \}
\end{aligned}
\label{eq:bim}
\end{equation}
where $\mathbf{x}$ is the input sample, $\mathbf{x}_i^*$ is the crafted adversarial sample (to be added to the input of the victim) at $i$\textsuperscript{th} iteration, $\bm{\theta}$ is the vector containing AI model weights, $J$ is the objective function, $\epsilon$ is a tunable parameter, limiting maximum level of perturbation for $L_{\infty}$ norm, $\alpha$ is the step size, and $\text{Clip}_{\epsilon}\{\cdot\}$ is the clipping operator that clips entries of the argument larger than $\epsilon$ to $\epsilon$ and smaller than -$\epsilon$ to -$\epsilon$.

\subsection{Sources of Adversarial Attack Threats in D-MIMO}

\begin{figure}[ht]
    \centering
    \includegraphics[width=0.35\textwidth]{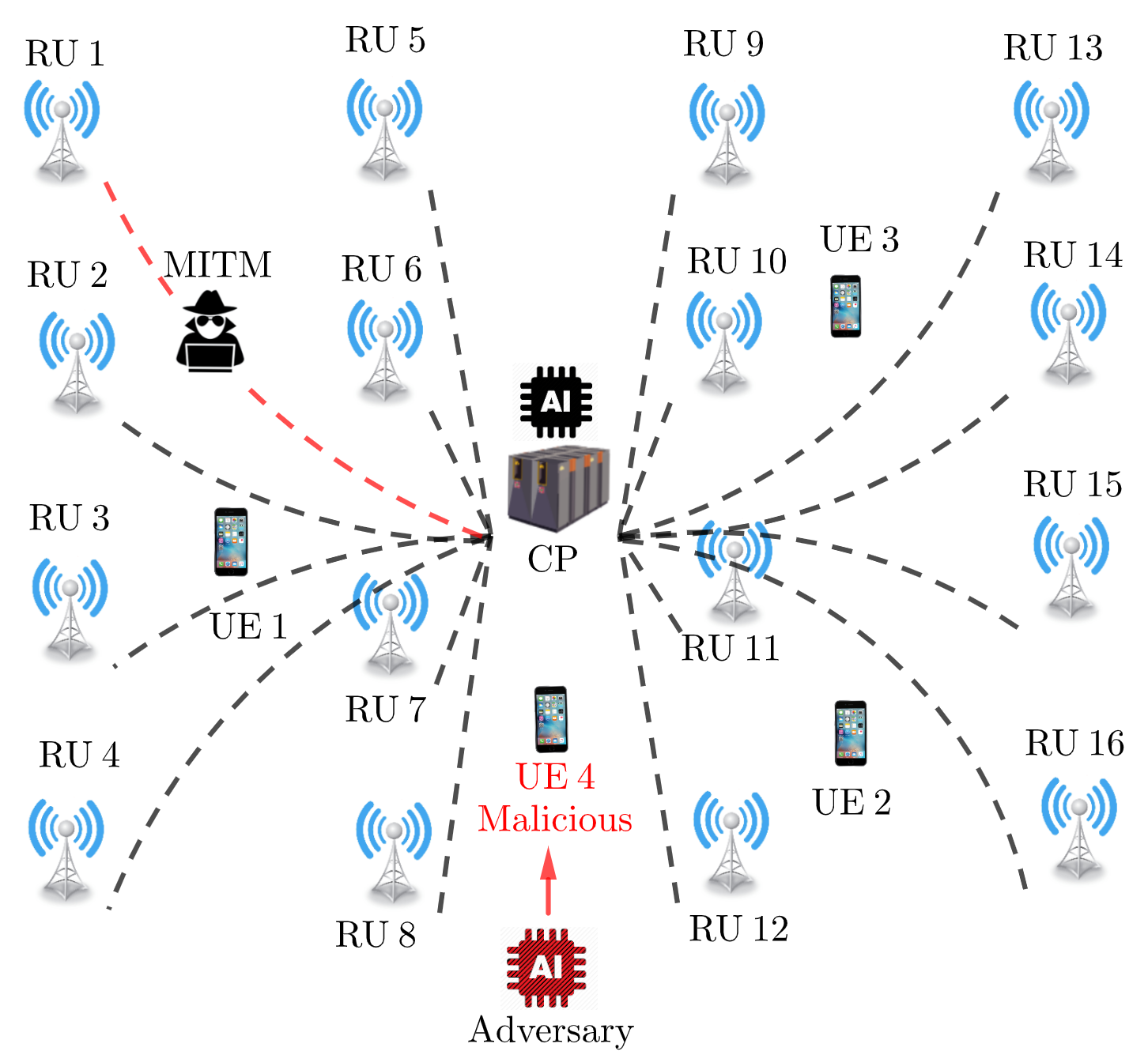}
    \caption{D-MIMO network with potential attacks.}
    \label{fig2}
    \vspace{-5mm}
\end{figure}

In Fig. \ref{fig2}, we demonstrate a D-MIMO network with potential attack sources. The first possible attack scenario is related to malicious RUs or fronthaul links. A man-in-the-middle (MITM) attack can be applied to change the channel related data transmitted from RUs to CP by an adversary. This type of attack enables to perturb the channel data of RUs on which the attack is applied. The second scenario is related to the UEs. There may be some malicious UEs in the network with modified RF and baseband components so that the corresponding channel information of those UEs are perturbed via applying some perturbation to pilot signals. In TDD systems, the channel knowledge is obtained at RU side using uplink pilots received from UEs. Therefore, a change in pilot signalling can produce false channel knowledge at RU side. We will analyze the effects of both these attack types in section \ref{simulation_result}.
\subsection{Practical Limitations of the Adversary}
From the adversary's perspective, there are three important constraints which limits the success of the adversarial attack in a D-MIMO network. Firstly, the adversary mostly does not have access to the details (architecture and weights) of the original AI model, therefore cannot use it in a white-box setting for crafting adversarial samples. Secondly, the adversary may not have complete knowledge of the input features of the AI model. Because, it is almost impossible for the adversary to know the channel information of each UE. Lastly, in a practical scenario, the adversary does not have the capability to introduce perturbations to all parts of the input vector, even if the channel information is known beforehand. However, despite all these limitations, there are proven ways in literature which increase the success of the attacker. Regarding the first limitation, it has been shown that a surrogate AI model might be sufficient to launch an effective attack \cite{transferability} due to the transferability nature of the adversarial samples. Regarding the other limitation, the universal adversarial perturbation (UAP) method \cite{uap} is proposed for cases where the complete input knowledge is not available.

\subsection{Proposed modified UAP (m-UAP) Method} 

To craft a perturbation when there is partial input knowledge, we propose a modified UAP (m-UAP) method which is based on the version suggested by Santos et al. \cite{santos} as in Algorithm 1 where principal component analysis (PCA) is applied to a set of output vectors obtained by the attack method.

\begin{algorithm}[!ht]
\DontPrintSemicolon
  \KwInput{$\textbf{x}, \epsilon$}
  \KwData{$\{\textbf{x}_1, \textbf{x}_2, \ldots, \textbf{x}_N\}, \widetilde{f}(\cdot,\bm{\theta})$}
  \KwOutput{$\bm{\delta}$}
  Define a matrix $\widehat{\textbf{X}}^{N \times MK}$ using the known entries of the input $\textbf{x}$ and the vectors $\{\textbf{x}_1, \textbf{x}_2, \ldots, \textbf{x}_N\}$ taken from the test set. Firstly, initialize the matrix $\widehat{\textbf{X}}$ as $\widehat{\textbf{X}}=[\textbf{x}_1 \: \textbf{x}_2 \: \cdots \textbf{x}_N]^T$. Then update $\widehat{\textbf{X}}$ using the known entries of $\textbf{x}$, i.e., $[\widehat{\textbf{X}}]_{i,j}=[\textbf{x}]_j$ for all known term indices $j$ and for all $i=1, 2, \ldots, N$.  \\
  For each row $\widehat{\textbf{x}}_i^T$ of $\widehat{\textbf{X}}$, apply the BIM to generate the matrix $\bm{P}^{N \times MK}=[\bm{\rho}_{\widehat{\textbf{x}}_1}, \bm{\rho}_{\widehat{\textbf{x}}_2}, \ldots, \bm{\rho}_{\widehat{\textbf{x}}_N}]^T=[\nabla_{\widehat{\textbf{x}}_1}J (\widetilde{f}(\widehat{\textbf{x}}_1, \bm{\theta})), \nabla_{\widehat{\textbf{x}}_2}J (\widetilde{f}(\widehat{\textbf{x}}_2, \bm{\theta})), \ldots, \nabla_{\widehat{\textbf{x}}_N}J (\widetilde{f}(\widehat{\textbf{x}}_N, \bm{\theta}))]^T$ \\
  Compute the principal right singular vector $\textbf{v}_1$ of $\textbf{X}$ as $\bm{P}=\textbf{U}\bm{\Sigma}\textbf{V}^H$, and $\textbf{v}_1$ is the first column of $\textbf{V}$. \\
  Compute two perturbations $\bm{\delta}_1=\epsilon\cdot \text{sign}(\textbf{v}_1), \: \bm{\delta}_2=-\bm{\delta}_1$ and calculate the sum objectives corresponding to the perturbed inputs, i.e.,  $J_u=\displaystyle\sum_{i=1}^N J(\widehat{\textbf{x}}_i+\bm{\delta}_u, \bm{\theta})$ for $u=1, 2$. Find the index $u_0 \in \{1, 2\}$ such that $u_0 = \underset{u \in \{1, 2\}}{\text{argmax}} \: J_u$. \\
  $\bm{\delta}=\bm{\delta}_{u_0}$. \\
  \Return $\bm{\delta}$
\caption{PCA-based modified UAP (m-UAP) method under $L_{\infty}$ norm}
\end{algorithm}

In our algorithm, $\widetilde{f}(\cdot, \bm{\theta})$ is the surrogate AI model used by the adversary, $\epsilon$ is the maximum allowed perturbation amount per each entry, $J$ is the objective function which is defined as the sum SE of users, i.e., $J(\textbf{x}, \bm{\theta}) = \displaystyle\sum_{k=1}^K \text{SE}_k$, and $\{\textbf{x}_1, \textbf{x}_2, \ldots, \textbf{x}_N\}$ are the $N$ randomly selected input samples from the test data set. We generate a matrix $\widehat{\textbf{X}}$ using the known entries of the input vector $\textbf{x}$ and other unknown elements are selected from $\textbf{x}_i$'s. Because, in a practical scenario, the adversary will most likely has access to the AI input data of some UEs or RUs only. Therefore, some part of the AI model's input is already known and modifiable by the adversary. After the generation of the matrix $\widehat{\textbf{X}}$, we apply BIM to each row to obtain the accumulated gradient ($\bm{\rho}_{\widehat{\textbf{x}}_i}$) of the objective function for all $i=1, 2, \ldots, N$. The BIM results are collected in a matrix $\textbf{P}$ and PCA is applied to find the principal right singular vector $\textbf{v}_1$. By this, we are able to obtain a generalized perturbation direction that reflects $\bm{\rho}_{\widehat{\textbf{x}}_i}$'s common characteristics. As $-\textbf{v}_1$ is also a valid right singular vector corresponding to the same singular value, we check which one to choose by evaluating the corresponding sum SEs of users. Here we evaluate the SE values using the input vectors obtained by the rows of $\widehat{\textbf{X}}$. Perturbations are calculated using the direction of the corresponding singular vectors and the perturbation limit $\epsilon$ to limit $L_{\infty}$ norm of the perturbation vector.

Lastly, it is important to choose the right objective function in a practical manner. In our simulations, our objective was to minimize the sum SE of UEs. If we instead chose to maximize sum of allocated powers above a predetermined limit as in \cite{santos}, then the attack would easily be mitigated by the network by normalizing the estimated power values.

\section{Simulation Results}\label{simulation_result}

\subsection{Experimental Setup}

We setup a simulation environment in MATLAB to generate our experimental training and test data sets to be used in model training and attack scenarios. We assume that there are 16 RUs distributed on a uniform grid in a $500 \text{m} \times 500 \text{m}$ area. There are 4 UEs and their locations are randomly chosen in the same region at each trial. All RU-UE channels have some large and small-scale fading parts, and they are assumed to be independent and constant within a coherence block. Large-scale part consists of path-loss with a 3-slope model and log-normal shadowing with a standard deviation $8$ dB, and small-scale part is assumed to be Rayleigh. We take $P_t=0.2$ W, $\sigma_k^2=-92$ dBm and $\text{BW}_k=20$ MHz for all $k$. These parameters are taken from \cite{ngo2017cell} and \cite{nayebi2017precoding}. We generate 400K pair of data consisting of large-scale fading coefficients ($\beta_{m,k}$'s) and their associated optimum power control coefficients ($\eta_{m,k}$'s). We use 390K pairs for model training and 10K pair of data for our tests.

\begin{figure*}[t!]
\centering
\includegraphics[width=0.9\textwidth]{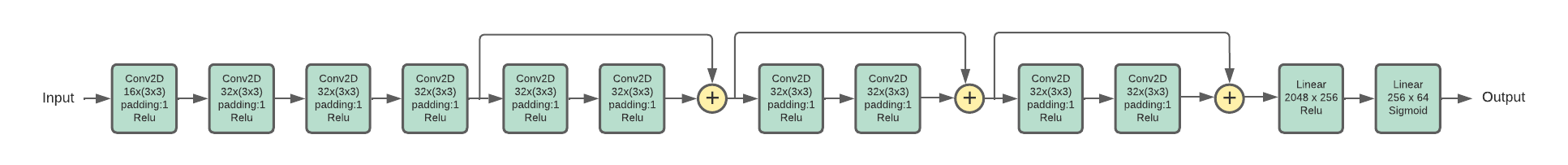}
\caption{Original AI Model}
\label{fig3}
\vspace{-5mm}
\end{figure*}

\begin{figure*}[t!]
\centering
\includegraphics[width=0.9\textwidth]{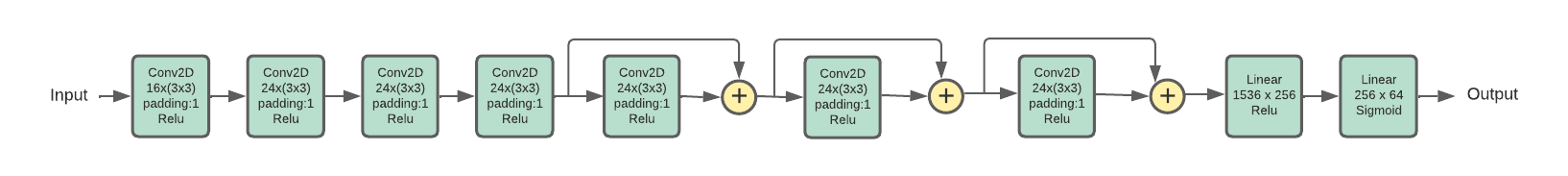}
\caption{Surrogate AI Model}
\label{fig4}
\vspace{-5mm}
\end{figure*}

\begin{figure}[ht]
    \centering
    \includegraphics[width=0.38\textwidth]{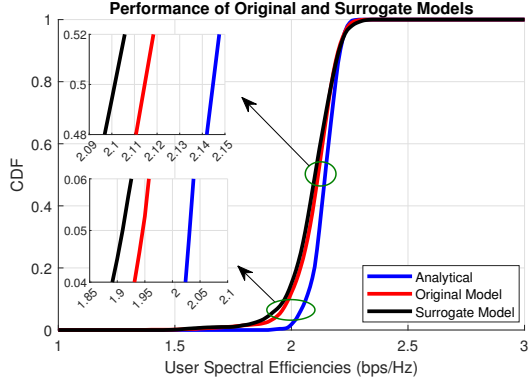}
    \caption{Performance of original and surrogate models.}
    \label{fig5}
    \vspace{-2mm}
\end{figure}

We use DNN-based regression models to learn the mapping between the large-scale fading coefficient vector $\textbf{x}=[\beta_{1,1} \: \beta_{1,2} \: \ldots \: \beta_{M,K}]^T \in\mathbb{R}^{MK}$ and the power control coefficient vector $\bm{\eta}=[\eta_{1,1} \: \eta_{1,2} \: \ldots \: \eta_{M,K}]^T \in\mathbb{R}^{MK}$. We train two different AI models which we denote as $f$ and $\widetilde{f}$ for the original AI model used in D-MIMO system and the surrogate model that the adversary uses to craft adversarial samples, respectively. Detailed model architectures are given in Fig. \ref{fig3} and \ref{fig4}. In Fig. \ref{fig5}, we present the per-user CDFs for the original and surrogate AI models, together with the result of analytical solution. We observe that surrogate model performance is slightly worse than the original one. Furthermore, both models can accurately find an approximate MMF solution as there is a small gap between the CDF curves of AI models and the CDF curve of the analytical result.


\subsection{Experimental Results}
In this section, we perform several simulations to see the effects of adversarial attacks under several constraints. 

We assume that the original model $f$ is used in CP to determine the power allocation. However, we assume that the adversary mainly uses the surrogate model $\widetilde{f}$ to craft adversarial samples. To see the performance gap from adversary's point of view, we also present the results in the case where the adversary has access to the original model. In our simulations, we present the effects of the adversarial attacks on per-user SEs and average EE with different amounts of perturbations and different levels of input information.

\subsubsection{The ultimate performance of adversarial attacks}
\begin{figure}[ht]
    \centering
    \includegraphics[width=0.38\textwidth]{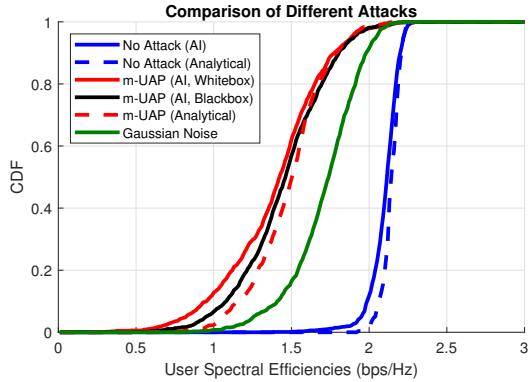}
    \caption{Comparison of different attack types ($\epsilon = 8$ dB).}
    \label{fig6}
    \vspace{-5mm}
\end{figure}

We begin our simulations by firstly showing the extreme scenario with the most devastating consequence where the adversary has access (read/modify) to all the parts of input vector fed to the AI model in CP. We set $\epsilon = 8$ dB which is equal to the standard deviation of the shadowing in the system. In Fig. \ref{fig6}, we see the CDFs of per-user SEs under different attack types. 
It is clear that m-UAP attack results in much more devastating consequences than standard Gaussian perturbation. Under the attack, the performance of the analytical solution also degrades showing that adversarial training is not suitable for this regression task. Because applying adversarial training will degrade the natural (clean) performance of the system. This shows the necessity of developing new defense solutions for such kind of smart attack threats. Finally, we observe that surrogate model (indicated as blackbox) has slightly worse disruptive performance than the original model (indicated as whitebox). This result proves that the adversary does not need to have access to the original AI model for crafting effective adversarial samples. 

\subsubsection{The performance with partial input knowledge}

To cover more practical cases which might be seen in a real-world scenario, we assume that the adversary might have a partial channel knowledge and partial perturbation capability on the input. In Fig. 7-8, we present the median (the SE value corresponding to $\text{CDF} = 0.5$) and the 5th percentile (the SE value corresponding to $\text{CDF} = 0.05$) per-user SEs for various levels of input information and perturbation capability. Fig. 7 shows the results of the first type of attack scenario which might be originated by malicious RUs. Fig. 8 shows the results of the second type of attack which might be launched by malicious UEs. In each of these cases, we assume that some portion of the UEs or RUs are malicious and only their channel information can be known and perturbed by the adversary. In both scenarios, the adversary uses the surrogate model with $\epsilon = 8$ dB.

\begin{figure}[ht]
    \centering
    \includegraphics[width=0.37\textwidth]{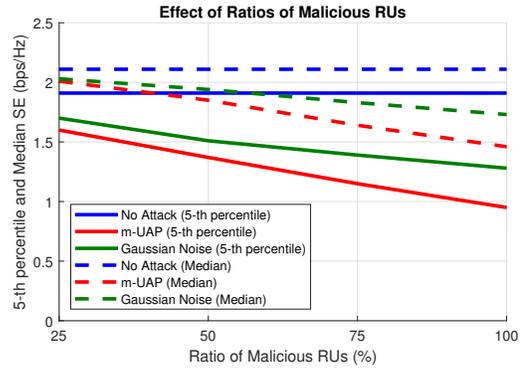}
    \caption{The effect of attack on RUs ($\epsilon = 8$ dB).}
    \label{fig7}
    \vspace{-2mm}
\end{figure}
\vspace{-3mm}
\begin{figure}[ht]
    \centering
    \includegraphics[width=0.37\textwidth]{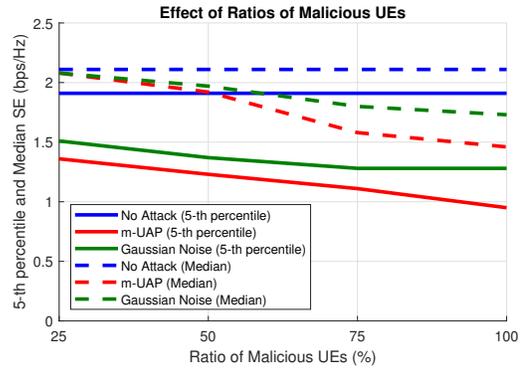}
    \caption{The effect of attack on UEs ($\epsilon = 8$ dB).}
    \label{fig8}
    \vspace{-2mm}
\end{figure}

According to the results obtained in Fig. 7-8, we conclude that m-UAP attack outperforms conventional approach of applying standard Gaussian noise in all cases. The gap gets larger when the ratio of involved malicious actors in the network increases. When the level of information and perturbation capability about the input become larger, the adversary can degrade the performance more, as expected. In all these adversarial attack cases, we observe a significant decrease in the user SE performance of the D-MIMO network.

\subsubsection{The effect of amount of perturbation}

The results obtained so far assumes $\epsilon=8$ dB. Considering the shadowing standard deviation (which is equal to $8$ dB), it is theoretically very hard for the system to detect an attack with $\epsilon = 8$ dB as the probability of observing that much variation in large-scale fading coefficients is roughly $32\%$, which is not negligible. On the other hand, for $\epsilon>16$ dB, the same probability decreases down to $5\%$ making the detection possible. 

In Fig. 9, we observe the effect of $\epsilon$ for values up to $16$ dB. We consider the scenario where half of the RUs are malicious and employ m-UAP method using a surrogate model.

\begin{figure}[ht]
    \centering
    \includegraphics[width=0.36\textwidth]{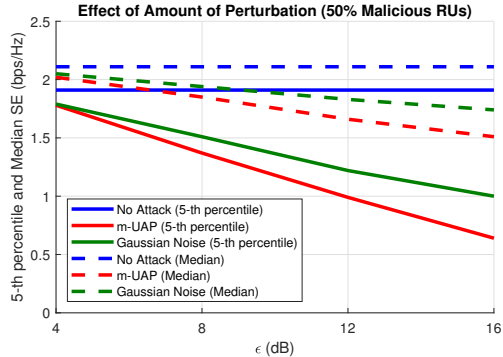}
    \caption{The effect of different $\epsilon$ values.}
    \label{fig9}
    \vspace{-1mm}
\end{figure}

The results show that adversarial attacks are more effective than standard Gaussian noise, and the attack becomes more disruptive as $\epsilon$ increases. 
\subsubsection{The effect on energy efficiency}

Lastly, we consider the impacts of the adversarial attacks on the EE in the network as sustainability is one of the important concerns in 6G. The power consumption may differ under attack, so we need to analyze the EE separately. The average EE for various $\epsilon$ are analyzed for the case where half of the RUs are malicious and we employ m-UAP method using a surrogate model.

\begin{figure}[ht]
    \centering
    \includegraphics[width=0.37\textwidth]{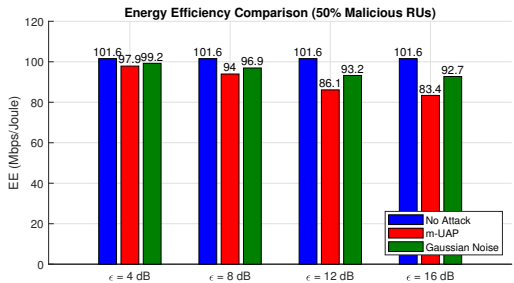}
    \caption{The effect on energy efficiency for various $\epsilon$ values.}
    \label{fig10}
    \vspace{-1mm}
\end{figure}

In Fig. 10, we see the comparison of no attack, adversarial attack and standard Gaussian noise attack cases. The adversary can decrease EE more than $8$ percent in $\epsilon=8$ dB case and the effect becomes more severe with increasing $\epsilon$ values. We conclude that adversarial attacks can also degrade the energy efficiency of the system.

\section{Conclusion}
In this study, we investigate the potential effects of adversarial attacks on AI-driven power control in D-MIMO. We propose m-UAP method to craft input specific yet PCA-based perturbations by making use of the partial knowledge about the channel. We consider different practical attack types and analyze the disruptive impacts of adversarial attacks under various levels of knowledge and capabilities. The results show that adversarial attacks with optimized perturbations might have a potential to degrade the performance of the network in terms of both spectral and energy efficiency. Smart defense techniques should not be neglected when deploying D-MIMO networks.

\section*{Acknowledgment}
This work was supported by The Scientific and Technological
Research Council of Turkey (TUBITAK) through the 1515
Frontier Research and Development Laboratories Support Program
under Project 5169902, and has been partly funded by
the European Commission through the H2020 project Hexa-X
(Grant Agreement no. 101015956).

\bibliographystyle{IEEEtran}
\bibliography{references}

\end{document}